\documentclass[a4paper,12pt]{article}

\newcommand{\be}{\begin{equation}}
\newcommand{\ee}{\end{equation}}

\newcommand{\spose}[1]{\hbox to 0pt{#1\hss}}
\newcommand{\lta}{\mathrel{\spose{\lower 3pt\hbox{$\mathchar"218$}}
 \raise 2.0pt\hbox{$\mathchar"13C$}}}
\newcommand{\gta}{\mathrel{\spose{\lower 3pt\hbox{$\mathchar"218$}}
 \raise 2.0pt\hbox{$\mathchar"13E$}}}

\begin{document}

\author{Micha{\l} Chodorowski}

\title{\vspace{1cm} Eppur si muove}
\date{\small \it Copernicus Astronomical Center, Bartycka 18, 00--716 
Warsaw, Poland }
\maketitle

\begin{abstract}
\normalsize In two recent papers, Abramowicz et al.~claim that the
expansion of the Universe can be interpreted only as the expansion of
space. In fact, what they really prove is that the cosmological
expansion cannot be described in terms of real motions in {\em
Minkowski} spacetime. However, there is no controversy about this
issue. Abramowicz et al.~show that in general, the cosmological redshift
is not a Doppler shift and they consider this fact as a proof that space
expands. Again, nobody believes (perhaps except Milne) that for
non-empty universes the origin of the redshift is purely
Dopplerian. From the Principle of Equivalence it follows that there must
be also a gravitational shift in presence of matter. Indeed, it is well
known in cosmology that for small redshifts, the cosmological redshift
can be decomposed into a Doppler component and a gravitational
component. In a forthcoming paper, we shall perform such a decomposition
for arbitrarily large values of the redshift.
\end{abstract}

\section{Introduction}
\label{sec:intro} Erroneous identification of Special Relativity (SR)
with `real motions', and General Relativity (GR) with the `expansion of
space' has a long history in cosmology. This misconception dates back to
Milne (1933) (see Chodorowski 2007 for a discussion).  Unfortunately, it
has been inherited and is shared by many contemporary authors, including
Abramowicz et al.~(2007, 2008). They show that Friedman-Lema{\^\i}tre
cosmological models are not (except for the empty model) compatible with
SR, and use this fact as an argument for the expansion of space. Many
other facts and gedanken experiments have been presented as a proof for
the expansion of space, but thus far, they have been all abolished (see
Chodorowski 2007 for a description and references; Lewis et al.~2008).

Obviously, defenders of the idea of real motions have a hard time,
because, as pointed out by Popper, you cannot prove a theory; you can
only falsify it. However, there is a long (and good) tradition in
science that faith in a given theory is proportional to the number of
failed attempts to disprove it. We hope that this will be also the case
for the idea of real motions. Personally, however, we have no longer
patience to be actively involved in the subject. Let others do
it. Indeed, there are recent papers on this issue (Lewis et al.~2008,
Bunn and Hogg 2008), with similar conclusions to ours.

Bunn and Hogg show that in `some sense' the cosmological redshift is a
Doppler shift. We don't think that this particular `sense' is the best
one, but their paper suggests an interesting way of decomposing the
cosmological redshift into a Doppler component and a gravitational
component, for any value of the redshift. It is well known how to do
this for small redshifts (e.g.~Bondi 1947, Peacock 1999), but not for
large redshifts. There have been some attempts to do this in the past
(Infeld and Schild 1945, Fock 1955), but for reasons we will describe
elsewhere, they were not successful. In a forthcoming paper, using fully
generally-relativistic approach we perform such a decomposition for
arbitrary redshifts. In the limit of small redshifts, our result reduces
to the well-known second-order formula.

Our Universe is in fact inhomogeneous: fluctuations of matter and
radiation are present (though small) even on the largest scales. How to
describe this fact in terms of expanding space? Does space expand at a
different rate in different points? Still, the concept of expanding
space seems to be a useful teaching aid to visualize uniform
expansion. You take a balloon. You draw dots on its surface. You tell
students that the dots represent galaxies. You blow up the balloon.
Distances between points increase. Students are fully convinced that
galaxies really remain in rest (by the way, relative to what?) and this
is the expansion of space that causes them to separate. Simple?
Simple. But in reality, the balloon does not exist; there are only
dots. GR is a classical theory. As such, it describes motions of
relativistically moving massive bodies in presence of gravity. The
concept of spacetime is central in this theory, but endowing space with
physical properties is incorrect.

Finally, let us quote Steven Weinberg (1993). `(...) space does not
expand. Cosmologists sometimes talk about expanding space -- but they
should know better.'

\end{document}